\RequirePackage{lineno}
\documentclass[aps,prl,twocolumn,showpacs,superscriptaddress,groupedaddress]{revtex4}  
\usepackage{graphicx}  
\usepackage{dcolumn}   
\usepackage{bm}        
\usepackage{amssymb}   
\usepackage{multirow}
\usepackage{epsfig}
\usepackage{amsmath}

\newcommand{\DY}{p\bar{p}\rightarrow Z/\gamma^{*}\rightarrow e^{+}e^{-}}
\newcommand{\effstw}{\ensuremath{\sin^2\theta_{\text{eff}}^{\text{$\ell$}}}}

\newcommand{\stw}{\ensuremath{\sin^2 \theta_{W}}}

\begin{document}


\modulolinenumbers[3]
\lefthyphenmin=2
\righthyphenmin=2

\widetext

\hspace{5.2in} \mbox{FERMILAB-PUB-14-288-E}
\title{Measurement of the effective weak mixing angle in
$\boldsymbol{\DY}$ events}

%
\affiliation{LAFEX, Centro Brasileiro de Pesquisas F\'{i}sicas, Rio de Janeiro, Brazil}
\affiliation{Universidade do Estado do Rio de Janeiro, Rio de Janeiro, Brazil}
\affiliation{Universidade Federal do ABC, Santo Andr\'e, Brazil}
\affiliation{University of Science and Technology of China, Hefei, People's Republic of China}
\affiliation{Universidad de los Andes, Bogot\'a, Colombia}
\affiliation{Charles University, Faculty of Mathematics and Physics, Center for Particle Physics, Prague, Czech Republic}
\affiliation{Czech Technical University in Prague, Prague, Czech Republic}
\affiliation{Institute of Physics, Academy of Sciences of the Czech Republic, Prague, Czech Republic}
\affiliation{Universidad San Francisco de Quito, Quito, Ecuador}
\affiliation{LPC, Universit\'e Blaise Pascal, CNRS/IN2P3, Clermont, France}
\affiliation{LPSC, Universit\'e Joseph Fourier Grenoble 1, CNRS/IN2P3, Institut National Polytechnique de Grenoble, Grenoble, France}
\affiliation{CPPM, Aix-Marseille Universit\'e, CNRS/IN2P3, Marseille, France}
\affiliation{LAL, Universit\'e Paris-Sud, CNRS/IN2P3, Orsay, France}
\affiliation{LPNHE, Universit\'es Paris VI and VII, CNRS/IN2P3, Paris, France}
\affiliation{CEA, Irfu, SPP, Saclay, France}
\affiliation{IPHC, Universit\'e de Strasbourg, CNRS/IN2P3, Strasbourg, France}
\affiliation{IPNL, Universit\'e Lyon 1, CNRS/IN2P3, Villeurbanne, France and Universit\'e de Lyon, Lyon, France}
\affiliation{III. Physikalisches Institut A, RWTH Aachen University, Aachen, Germany}
\affiliation{Physikalisches Institut, Universit\"at Freiburg, Freiburg, Germany}
\affiliation{II. Physikalisches Institut, Georg-August-Universit\"at G\"ottingen, G\"ottingen, Germany}
\affiliation{Institut f\"ur Physik, Universit\"at Mainz, Mainz, Germany}
\affiliation{Ludwig-Maximilians-Universit\"at M\"unchen, M\"unchen, Germany}
\affiliation{Panjab University, Chandigarh, India}
\affiliation{Delhi University, Delhi, India}
\affiliation{Tata Institute of Fundamental Research, Mumbai, India}
\affiliation{University College Dublin, Dublin, Ireland}
\affiliation{Korea Detector Laboratory, Korea University, Seoul, Korea}
\affiliation{CINVESTAV, Mexico City, Mexico}
\affiliation{Nikhef, Science Park, Amsterdam, the Netherlands}
\affiliation{Radboud University Nijmegen, Nijmegen, the Netherlands}
\affiliation{Joint Institute for Nuclear Research, Dubna, Russia}
\affiliation{Institute for Theoretical and Experimental Physics, Moscow, Russia}
\affiliation{Moscow State University, Moscow, Russia}
\affiliation{Institute for High Energy Physics, Protvino, Russia}
\affiliation{Petersburg Nuclear Physics Institute, St. Petersburg, Russia}
\affiliation{Instituci\'{o} Catalana de Recerca i Estudis Avan\c{c}ats (ICREA) and Institut de F\'{i}sica d'Altes Energies (IFAE), Barcelona, Spain}
\affiliation{Uppsala University, Uppsala, Sweden}
\affiliation{Taras Shevchenko National University of Kyiv, Kiev, Ukraine}
\affiliation{Lancaster University, Lancaster LA1 4YB, United Kingdom}
\affiliation{Imperial College London, London SW7 2AZ, United Kingdom}
\affiliation{The University of Manchester, Manchester M13 9PL, United Kingdom}
\affiliation{University of Arizona, Tucson, Arizona 85721, USA}
\affiliation{University of California Riverside, Riverside, California 92521, USA}
\affiliation{Florida State University, Tallahassee, Florida 32306, USA}
\affiliation{Fermi National Accelerator Laboratory, Batavia, Illinois 60510, USA}
\affiliation{University of Illinois at Chicago, Chicago, Illinois 60607, USA}
\affiliation{Northern Illinois University, DeKalb, Illinois 60115, USA}
\affiliation{Northwestern University, Evanston, Illinois 60208, USA}
\affiliation{Indiana University, Bloomington, Indiana 47405, USA}
\affiliation{Purdue University Calumet, Hammond, Indiana 46323, USA}
\affiliation{University of Notre Dame, Notre Dame, Indiana 46556, USA}
\affiliation{Iowa State University, Ames, Iowa 50011, USA}
\affiliation{University of Kansas, Lawrence, Kansas 66045, USA}
\affiliation{Louisiana Tech University, Ruston, Louisiana 71272, USA}
\affiliation{Northeastern University, Boston, Massachusetts 02115, USA}
\affiliation{University of Michigan, Ann Arbor, Michigan 48109, USA}
\affiliation{Michigan State University, East Lansing, Michigan 48824, USA}
\affiliation{University of Mississippi, University, Mississippi 38677, USA}
\affiliation{University of Nebraska, Lincoln, Nebraska 68588, USA}
\affiliation{Rutgers University, Piscataway, New Jersey 08855, USA}
\affiliation{Princeton University, Princeton, New Jersey 08544, USA}
\affiliation{State University of New York, Buffalo, New York 14260, USA}
\affiliation{University of Rochester, Rochester, New York 14627, USA}
\affiliation{State University of New York, Stony Brook, New York 11794, USA}
\affiliation{Brookhaven National Laboratory, Upton, New York 11973, USA}
\affiliation{Langston University, Langston, Oklahoma 73050, USA}
\affiliation{University of Oklahoma, Norman, Oklahoma 73019, USA}
\affiliation{Oklahoma State University, Stillwater, Oklahoma 74078, USA}
\affiliation{Brown University, Providence, Rhode Island 02912, USA}
\affiliation{University of Texas, Arlington, Texas 76019, USA}
\affiliation{Southern Methodist University, Dallas, Texas 75275, USA}
\affiliation{Rice University, Houston, Texas 77005, USA}
\affiliation{University of Virginia, Charlottesville, Virginia 22904, USA}
\affiliation{University of Washington, Seattle, Washington 98195, USA}
\author{V.M.~Abazov} \affiliation{Joint Institute for Nuclear Research, Dubna, Russia}
\author{B.~Abbott} \affiliation{University of Oklahoma, Norman, Oklahoma 73019, USA}
\author{B.S.~Acharya} \affiliation{Tata Institute of Fundamental Research, Mumbai, India}
\author{M.~Adams} \affiliation{University of Illinois at Chicago, Chicago, Illinois 60607, USA}
\author{T.~Adams} \affiliation{Florida State University, Tallahassee, Florida 32306, USA}
\author{J.P.~Agnew} \affiliation{The University of Manchester, Manchester M13 9PL, United Kingdom}
\author{G.D.~Alexeev} \affiliation{Joint Institute for Nuclear Research, Dubna, Russia}
\author{G.~Alkhazov} \affiliation{Petersburg Nuclear Physics Institute, St. Petersburg, Russia}
\author{A.~Alton$^{a}$} \affiliation{University of Michigan, Ann Arbor, Michigan 48109, USA}
\author{A.~Askew} \affiliation{Florida State University, Tallahassee, Florida 32306, USA}
% ...
%
%
\collaboration{The D0 Collaboration\footnote{with visitors from
$^{a}$Augustana College, Sioux Falls, SD, USA,
$^{b}$The University of Liverpool, Liverpool, UK,
$^{c}$DESY, Hamburg, Germany,
$^{d}$Universidad Michoacana de San Nicolas de Hidalgo, Morelia, Mexico
$^{e}$SLAC, Menlo Park, CA, USA,
$^{f}$University College London, London, UK,
$^{g}$Centro de Investigacion en Computacion - IPN, Mexico City, Mexico,
$^{h}$Universidade Estadual Paulista, S\~ao Paulo, Brazil,
$^{i}$Karlsruher Institut f\"ur Technologie (KIT) - Steinbuch Centre for Computing (SCC),
D-76128 Karlsruhe, Germany,
$^{j}$Office of Science, U.S. Department of Energy, Washington, D.C. 20585, USA,
$^{k}$American Association for the Advancement of Science, Washington, D.C. 20005, USA,
$^{l}$Kiev Institute for Nuclear Research, Kiev, Ukraine,
$^{m}$University of Maryland, College Park, Maryland 20742, USA
and
$^{n}$European Orgnaization for Nuclear Research (CERN), Geneva, Switzerland
}} \noaffiliation
\vskip 0.25cm

\date{August 21, 2014}
\begin{abstract}
We present a measurement of the fundamental parameter of the standard model,
the weak mixing angle $\effstw$ which
determines the relative strength of weak and electromagnetic interactions,
in $\DY$ events at
a center of mass energy of 1.96 TeV, using data corresponding
to 9.7 fb$^{-1}$ of integrated luminosity collected by the D0 detector at the Fermilab Tevatron. The effective
weak mixing angle is extracted from the forward-backward charge
asymmetry as a function of the invariant mass around
the $Z$ boson pole.
The measured value of $\effstw=0.23147 \pm 0.00047$ is the most
precise measurement from light quark interactions to date, with a
precision close to the best LEP and SLD results.
\end{abstract}
\maketitle

The weak mixing angle $\sin^2\theta_W$ is one of the
fundamental parameters of the standard model (SM). It describes the
relative strength of the axial-vector couplings $g^f_A$ to the
vector couplings $g^f_V$ in neutral-current interactions of a $Z$ boson
to fermions $f$ with Lagrangian
\begin{equation}
\mathcal{L} = -i
\frac{g}{2\cos\theta_W}\bar{f}\gamma^{\mu}\left( g^{f}_{V} -
g^{f}_{A}\gamma_5 \right) fZ_{\mu},
\end{equation}
with $g^{f}_{V} = I^f_3 - 2Q_f \cdot \sin^2\theta_W, g^f_A = I^f_3$,
where $I^f_3$ and $Q_f$ are the weak isospin component and the
charge of the fermion. At tree level and in all orders of the
on-shell renormalization scheme, the weak mixing angle can be written in
terms of the $W$ and $Z$ boson masses as $\sin^2\theta_W =1 -
M^2_W/M^2_Z$. To include higher order electroweak radiative corrections,
effective weak mixing angles are defined as
\begin{equation}
\sin^2\theta^f_{\text{eff}}=\frac{1}{4|Q_f|}\left(1-\frac{g_V^f}{g_A^f}\right),
\end{equation}
for each fermion flavor.

It is customary to quote the charged lepton effective weak mixing angle $\effstw$,
determined by measurements of observables around the $Z$ boson pole.
There is tension between the two most precise measurements of $\effstw$, which are
$0.23221 \pm 0.00029$ from the combined LEP measurement using
the charge asymmetry $A_{FB}^{0,b}$ for $b$ quark production
and $0.23098 \pm 0.00026$ from the SLD measurement
of the $e^+e^-$ left-right polarization
asymmetry $A_{lr}$~\cite{LEP-SLD}.
An independent determination of the effective weak mixing angle is therefore an
important precision test of the SM electroweak breaking mechanism.

At the Tevatron, the mixing angle can be measured in the Drell-Yan
process $p\bar{p} \rightarrow Z/\gamma^{*} \rightarrow \ell^+\ell^-$,
through a forward-backward charge asymmetry in the distribution of the emission angle
$\theta^*$ of the negatively charged lepton momentum
relative to the incoming quark momentum, defined in the
Collins-Soper frame~\cite{cs-frame}. Events with $\cos \theta^*>0$
are classified as forward ($F$), and those with $\cos \theta^*<0$
as backward ($B$). The forward-backward charge asymmetry, $A_{FB}$, is defined by
\begin{equation}
 A_{FB} = \frac{N_{F}-N_{B}}{N_F+N_B},
\end{equation}
where $N_{F}$ and $N_{B}$ are the numbers of forward and backward
events. The asymmetry arises from the interference between vector and axial vector
coupling terms.

The asymmetry $A_{FB}$ can be measured as a function of the
invariant mass of the dilepton pair ($M_{ee}$).
The presence of both vector and axial-vector couplings of the $Z$ boson to fermions
gives the most significant variation of $A_{FB}$ in
vicinity of the $Z$ boson pole, which is sensitive to the effective weak
mixing angle.

Measurements of $\effstw$ have been reported previously by the CDF Collaboration
in the $Z\rightarrow e^{+}e^{-}$~\cite{cdf_stw, cdf_stw_2fb}
and $Z\rightarrow \mu^{+}\mu^{-}$~\cite{cdf_muon} channels,
and the D0 Collaboration in the $Z\rightarrow e^{+}e^{-}$ channel~\cite{d0_1fb, d0-zee-5fb}.
The angle $\effstw$ has also been measured at the LHC in $pp$
collisions by the CMS Collaboration in the $Z\rightarrow \mu^{+}\mu^{-}$ channel
at $\sqrt{s}=7$~TeV~\cite{CMS_stw}.

This letter reports a measurement of the effective weak mixing angle from the $A_{FB}$ distribution
using 9.7 fb$^{-1}$ of integrated luminosity collected with the D0 detector
at the Fermilab Tevatron collider.
The precision of the previous D0 measurement using 5 fb$^{-1}$ of data~\cite{d0-zee-5fb},
$\effstw = 0.2309 \pm 0.0008~(\text{stat.}) \pm 0.0006~(\text{syst.})$, was dominated by the statistical uncertainty
and the uncertainty on the electron energy scale.
The analysis of the full 9.7 fb$^{-1}$ data set presented here features an extended acceptance and a new
electron energy calibration method providing substantially improved accuracy.

\indent The D0 detector comprises a central tracking
system, a calorimeter and a muon system~\cite{d0-detector, d0-det-2, d0-det-3}. The central tracking
system consists of a silicon microstrip tracker and a scintillating central fiber tracker,
both located within a 1.9~T superconducting solenoidal magnet and optimized for tracking and vertexing
capabilities at detector pseudorapidities of $|\eta_{\text{det}}|<3$~\cite{d0_coordinate}.
Outside the solenoid, three liquid argon and uranium calorimeters provide coverage of
$|\eta_{\text{det}}|<3.5$ for electrons: the central calorimeter (CC)
for $|\eta_{\text{det}}|<1.1$, and two endcap calorimeters (EC) in the range
$1.5<|\eta_{\text{det}}|<3.5$. Gaps between the cryostats
create inefficient electron detection regions between $1.1<|\eta_{\text{det}}|<1.5$
that are excluded from the analysis.
The muon system outside the calorimeter consists of drift chambers and scintillators before
and after iron toroid magnets. The solenoid and toroid polarities are reversed every two weeks on average.

The data used in this analysis are collected by triggers requiring at least
two electromagnetic (EM) clusters reconstructed in the calorimeter.
The determination of their energies uses only the calorimeter information.
Each EM cluster is further required to be in the CC or
EC, with transverse momentum $p_T$ $>25$ GeV, and to have shower shapes consistent
with that of an electron.
For events with
both EM candidates in the CC region (CC-CC), each EM object must have a
spatially matched track reconstructed in the tracking system.
For events with one EM cluster in the CC and the other in the EC region (CC-EC),
only the CC candidate is required to have a matched track.
For events with both candidates in the EC calorimeter (EC-EC), at least one EM
object must have a matched track. All tracks must have $p_T>10$~GeV
and satisfy track quality criteria to maintain a low charge mis-identification probability.
For CC-CC events, the two EM candidates are required to have opposite charges.
For CC-EC events, the determination of ``forward" or ``backward" is made according to the
charge measured for the track-matched CC EM candidate,
whereas the charge of the EC higher quality matched track
is used for EC-EC events~\cite{EMID_nim}.

Events are further required to have a reconstructed dielectron invariant mass in the range
$75<M_{ee}<115$~GeV. A larger sample satisfying $60<M_{ee}<130$~GeV is used to
understand detector responses and to tune the Monte Carlo (MC) simulation.

To maximize the acceptance, previously ignored electrons reconstructed near the boundaries of CC calorimeter
modules~\cite{d0-detector} ($\phi$-mod boundary) are included.
The geometric acceptance is further extended compared with previous D0 results~\cite{d0-zee-5fb} from
$|\eta_{\text{det}}|<1.0$ to $|\eta_{\text{det}}|<1.1$ for the CC, and from
$1.5<|\eta_{\text{det}}|<2.5$ to $1.5<|\eta_{\text{det}}|<3.2$ for the EC.
In addition, EC-EC events, which were excluded due to their
poorer track reconstruction and calorimeter energy resolution, are now included.
The extensions in $\eta_{\text{det}}$ and $\phi$-mod acceptance give a $70\%$
increase in the number of events over what would be expected from the increase in integrated
luminosity. An additional $15\%$ increase is gained
from improvements in the track reconstruction algorithm.
The number of $Z\rightarrow e^{+}e^{-}$ candidate events in the data sample is 560,267 which includes
248,380 CC-CC events, 240,593 CC-EC events and 71,294 EC-EC events.

The MC Drell-Yan $Z/\gamma^{*} \rightarrow e^+e^-$ sample is generated by
using the D0 simulation
software, based on the leading-order {\sc pythia} generator~\cite{pythia} with the
NNPDF2.3~\cite{NNPDF} parton distribution functions (PDFs), followed by a
{\sc geant}-based simulation~\cite{geant} of the D0 detector.
The {\sc pythia} MC samples, with data events from random beam crossings overlaid,
are mainly used to understand the geometric acceptance,
and the energy scale and resolution of electrons in the calorimeter.

A new method of electron energy calibration is developed and applied
to both the data and MC, which significantly reduces the
systematic uncertainty due to the electron energy measurement.
The weak mixing angle, which is extracted from $A_{FB}$ as a function of
$M_{ee}$, depends strongly on the position of the peak value of $M_{ee}$.
Therefore, it is critical to have a precise electron energy measurement
and a consistent measured peak value of $M_{ee}$ from different regions of the
detector across various Tevatron running conditions.
In Ref.~\cite{d0-zee-5fb}, an overall scale factor was applied to
simulations to model the detector response for electron energy depositions,
where the scale factor is determined by comparing the
$M_{ee}$ spectrum in data and MC, yielding a large
uncertainty due to background estimation and detector resolution.
In this analysis, a new energy calibration method is applied to the
data and the MC separately.
The energy measurement depends not only on the $\eta_{\text{det}}$, but
also on the instantaneous luminosity~\cite{WmassPRD}.
 For CC electrons, an instantaneous
luminosity-dependent scale factor ($\alpha_L^{\text{CC}}$) and an
$\eta_{\text{det}}$-dependent scale factor ($\alpha_{\eta}^{\text{CC}}$)
are applied to the electron energy.
For EC electrons in addition to the scale factors $\alpha_L^{\text{EC}}$ and $\alpha_{\eta}^{\text{EC}}$,
an $\eta_{\text{det}}$-dependent offset $\beta_{\eta}^{\text{EC}}$ is introduced
to model the $\eta_{\text{det}}$ dependence of the electron energy.
All correction factors are
determined by scaling the peak of the $M_{ee}$ distribution as a
function of instantaneous luminosity and $\eta_{\text{det}}$ to be consistent with
the $Z$ boson mass measured by LEP ($M_Z=91.1875$~GeV)~\cite{LEP-SLD}. The CC
correction factors are tuned with the CC-CC events. To
remove one degree of freedom, $\beta_{\eta}^{\text{EC}}$ is expressed as a
function of $\alpha_{\eta}^{\text{EC}}$, and the relationship is measured
with the CC-EC events. The values of $\alpha_{\eta}^{\text{EC}}$ and
$\beta_{\eta}^{\text{EC}}(\alpha_{\eta}^{\text{EC}})$ are fitted using the EC-EC
events. After the calibration, the standard deviation $\delta M$ of the $M_{ee}$ peak
values in different $\eta_{\text{det}}$ regions is $\approx$ 20 MeV.
Various closure tests are performed
to check the validity of the calibration procedure. For example, an
extra $\eta_{\text{det}}$-dependent offset is applied to the
corrected energy and fixed by performing the
calibration again. The extra offset is found to be consistent with $\delta M$.
The ratio of $\delta M$ to $M_Z$ is propagated into the uncertainty of the $\effstw$
measurement to estimate the systematic
uncertainty arising from the energy calibration.

After the electron energy calibration, an additional electron energy
resolution smearing is derived and applied to the MC to achieve agreement with the
width of the $M_{ee}$ distribution in data.
For the CC $\phi$-mod boundary electrons, the resolution smearing is
modeled with a Crystal Ball function~\cite{crystalball}.
For other electrons, the smearing is modeled with a Gaussian function.

Additional corrections and reweightings are applied to the MC simulation to
improve the agreement with data. The scale factors of the electron
identification efficiency between the MC and the data are measured
using the tag-and-probe method~\cite{tag-probe} and applied to the MC distributions
as functions of $p_T$ and $\eta_{\text{det}}$.
The simulation is further corrected for higher-order effects not included in {\sc pythia}
by reweighting the MC events at the generator level in two dimensions
($p_{T}$ and rapidity $y$ of the $Z$ boson)
to match {\sc resbos}~\cite{resbos} predictions. In addition, next-to-next-to-leading order QCD
corrections are applied as a function of $M_Z$~\cite{resbos, NNLO_QCD}.
The distribution of the instantaneous luminosity and the $z$ coordinate of the $p\bar{p}$ collision
vertices are also weighted to match those in the data.
Since $A_{FB}$ is defined as a ratio of numbers of events, many small uncertainties
cancel out. Only the electron selection
efficiency scale factor in these additional corrections contributes significantly to the
final uncertainty.

The charge of the particle track matched to the EM cluster is used to determine if
the EM cluster is associated to an electron or positron and to classify the event as
forward or backward.
The charge misidentification probability is given by the number of $\DY$ events reconstructed
with same-sign as a proportion of the total number of $\DY$ events.
The probabilities are measured in data and MC.
The charge of electrons and positrons reconstructed
in the MC is randomly changed to match the misidentification probability in the data averaged
over $p_T$ spectrum of electrons.
In the CC region the average charge misidentification rate in data is about $0.3\%$, whereas
in the EC region it varies from 1\% at $|\eta_{\text{det}}|=1.5$ to 10\% at $|\eta_{\text{det}}|=3.0$.
The statistical uncertainty of the measured charge
misidentification rate is included as a systematic uncertainty.

The background is suppressed by the strict requirements
on the quality of the matched track. The main contribution is from
multijet events, in which jets are misreconstructed as electrons,
and is estimated from data.
The multijet production from the proton anti-proton
initial state produces jets, and hence fake electrons, nearly symmetrically with respect to the forward and backward
hemispheres~\cite{d0-zee-5fb, d0-ttbar-FB}.
Multijet events are selected by reversing some of the electron selection cuts to
study the differential distributions
of the multijet background, which are different from
the real multijet background that passes all the electron
selections. Therefore, a correction factor is applied as a function of electron $p_T$, given by the ratios of
the efficiencies for EM-like jets (which are
selected in a multijet-enriched data sample and pass all the electron selections)
and ``reverse-selected'' jets.
The normalization of the multijet background is determined by fitting the
sum of the $M_{ee}$ distributions of multijet events and the signal MC events to
the distribution from the selected data events.
The $W$+jets, $Z/\gamma^* \rightarrow \tau\tau$, di-boson ($WW$ and $WZ$), $\gamma \gamma$ and $t\overline{t}$
backgrounds are estimated using
the {\sc pythia} MC simulations.
At the $Z$ boson peak, the multijet background is $0.3\%$ and the sum of the
di-boson, $W$+jets, $Z/\gamma^* \rightarrow \tau\tau$, $\gamma \gamma$ and $t\overline{t}$
backgrounds is about $0.05\%$.
The $M_{ee}$ and $\cos\theta^{*}$ distributions of data and of the sum of signal MC and background
expectations are in good agreement with the SM predictions~\cite{supplement}.

\begin{figure}[!hbt]
\begin{center}
 \epsfig{scale=0.45, file=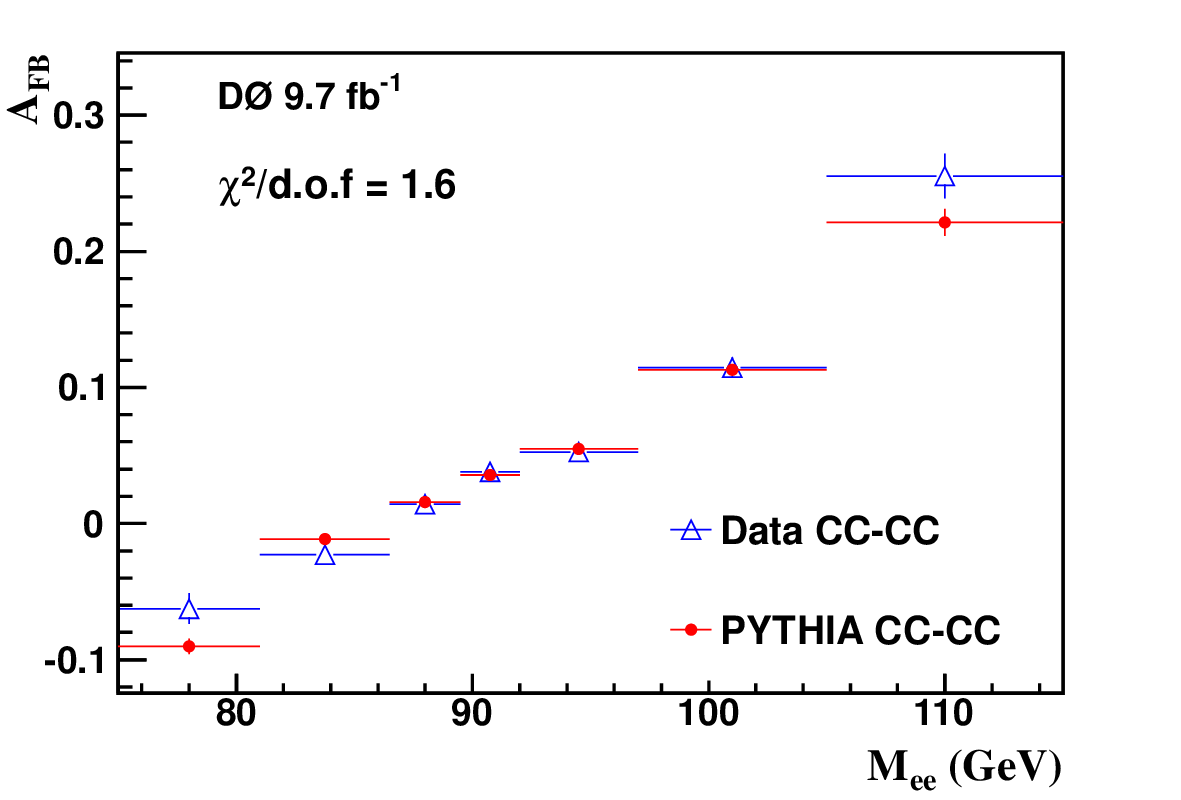}
 \epsfig{scale=0.45, file=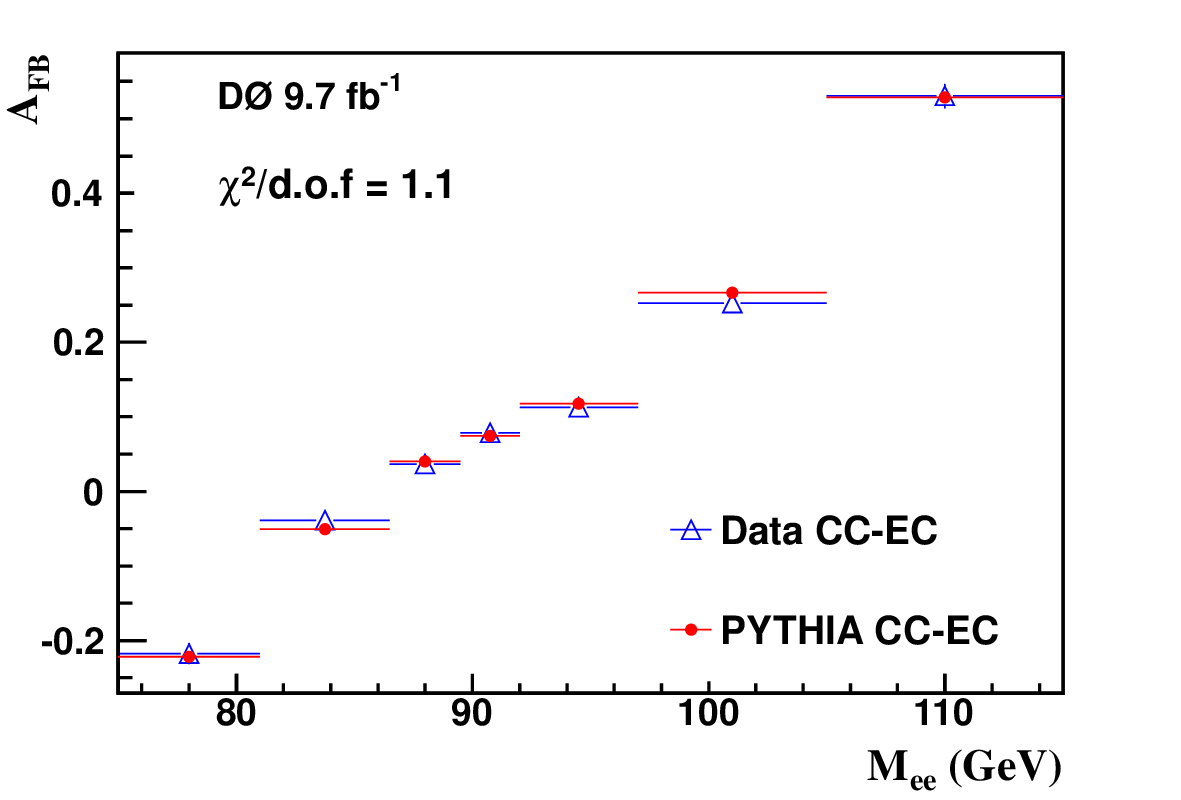}
 \epsfig{scale=0.45, file=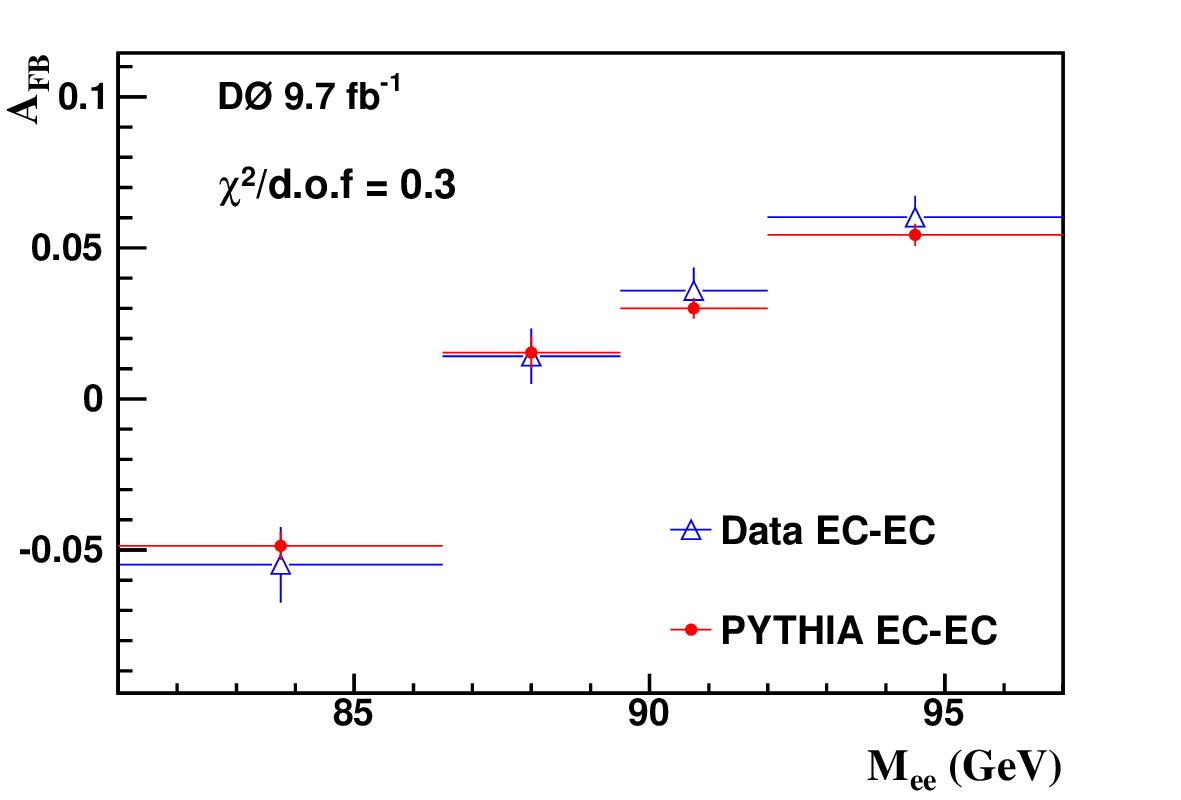}
 \caption{\small (color online). Comparison between the $A_{FB}$ distributions measured in
 the background-subtracted data and the MC for the three
 kinematic regions, with the corresponding $\chi^2$ per degree of freedom. $\stw$ in
 the MC is 0.23139. The error bars are statistical only.}
 \label{fig:rawAFB_cccc}
\end{center}
\end{figure}

The AFB distributions as a
function of mass are obtained for CC-CC, CC-EC and EC-EC categories by
summing the samples of specific solenoid and toroid magnet polarities,
after weighting each by the integrated luminosity for the sample. This
weighted combination provides cancellation of asymmetries due to
variations in detector response and acceptance with $\eta_{\text{det}}$ and $p_T$.
The weak mixing angle is extracted from the background-subtracted
$A_{FB}$ spectrum in the regions
$75<M_{ee}<115$~GeV for CC-CC and CC-EC events, and
$81<M_{ee}<97$~GeV for EC-EC events by comparing the data to simulated $A_{FB}$ templates
corresponding to different input values of $\stw$.
The mass window for EC-EC events is narrower to take into account the differences in track reconstruction
and energy measurement.
The templates are obtained by reweighting $M_Z$ and $\cos\theta^*$
distributions at the generator level to different Born-level
$\stw$ predictions. The $A_{FB}$ distribution is negligibly sensitive to
the effect of QED final state radiation because most of these radiated photons
are emitted co-linearly with the electron and are reconstructed as one single EM
object by the detector.
The background-subtracted $A_{FB}$
distribution and the {\sc pythia} prediction with the fitted $\stw$ are
shown in Fig.~\ref{fig:rawAFB_cccc}.

The results of the fits for different event categories,
with statistical and systematic uncertainties, are listed in Table~\ref{tab:result_iia}.
The uncertainties on $\stw$ are dominated by the limited
data sample. CC-EC events are the most sensitive to the weak mixing angle due to the larger
variation of $A_{FB}$ with mass in that kinematic region.
The systematic uncertainties due to the electron energy
calibration and resolution smearing, the estimation of the
backgrounds, the charge misidentification rate and the
identification efficiency are also dominated by the limited data sample.
We estimate the systematic uncertainty on the measured
$\stw$ due to instrumental asymmetries that remain after combining the
luminosity-weighted solenoid and toroid samples to be $\pm 0.00001$

The measurement is dominated by statistical uncertainties.
Systematic uncertainties are treated as uncorrelated but the total
uncertainty does not depend on whether they are taken to be correlated
or uncorrelated
The results were therefore combined by using the corresponding uncertainties as weights,
giving
\begin{eqnarray*}
 \lefteqn{\stw=}\\
 &&0.23139 \pm 0.00043\thinspace (\text{stat.}) \pm \\
 &&0.00008\thinspace (\text{syst.})\pm 0.00017\thinspace (\text{PDF}).
\end{eqnarray*}
The PDF uncertainty is estimated by reweighting the PDF set in the MC simulations
to different sets of the NNPDF2.3, computing the $\stw$ value for each set, and taking the
standard deviation of these values as the uncertainty~\cite{NNPDF}.

\begin{table}
\begin{center}
\begin{tabular}{l|cccc}
\hline \hline
   &  CC-CC  &  CC-EC  &  EC-EC & Combined\\
\hline
    $\stw$  &   0.23142  &   0.23143  &   0.22977  & 0.23139 \\ \hline
   Statistical  &  0.00116   &  0.00047   &   0.00276 & 0.00043 \\
\hline
   Systematic   &  0.00009  &  0.00009   &   0.00019 & 0.00008 \\
   ~~Energy Calibration      &  0.00003   &  0.00001   &   0.00004  & 0.00001 \\
   ~~Energy Smearing      &  0.00001   &  0.00002   &   0.00013 & 0.00002 \\
   ~~Background       &  0.00002  &  0.00001   &   0.00002 & 0.00001 \\
   ~~Charge Misidentification      &  0.00002   &  0.00004   &   0.00012 & 0.00003 \\
   ~~Electron Identification        &  0.00008   &  0.00008   &   0.00005 & 0.00007 \\
   ~~Fiducial Asymmetry    & 0.00002 & 0.00001 & 0.00001 & 0.00001 \\
\hline
   Total   & 0.00116 & 0.00048 & 0.00277 & 0.00044 \\
\hline \hline
\end{tabular}
\caption{\small Measured $\stw$ values and corresponding uncertainties.
Uncertainties from higher-order corrections and the PDFs are not included.}
\label{tab:result_iia}
\end{center}
\end{table}

\begin{figure}[!hbt]
\begin{center}
 \epsfig{scale=0.45, file=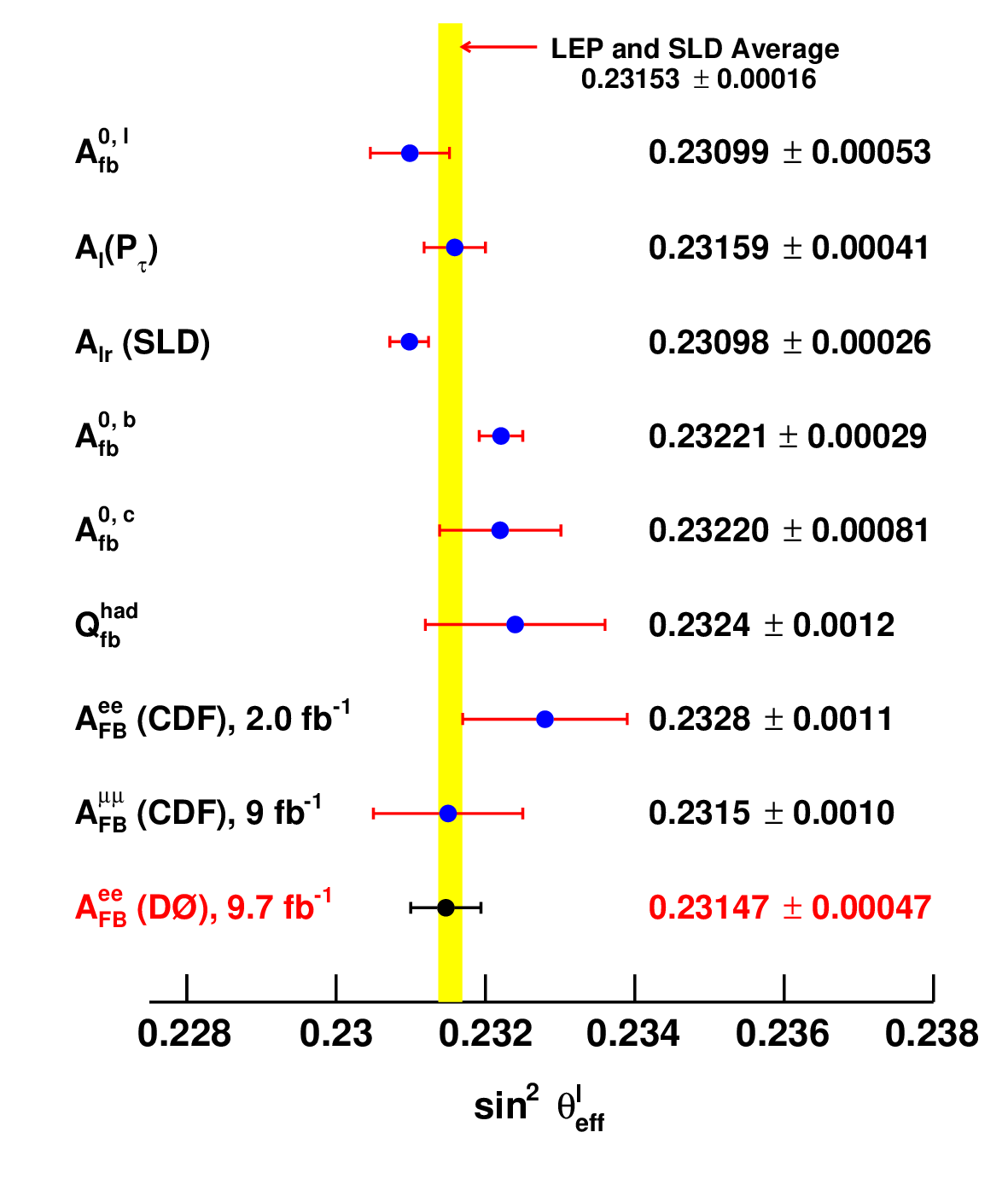}
\caption{\small (color online). Comparison of measured $\effstw$ with
results from other experiments.
The average is a combination of $A_{FB}^{0,\ell}$, $A_{l}(P_{\tau})$,
$A_{lr}(\text{SLD})$, $A_{FB}^{0,b}$, $A_{FB}^{0,c}$, and
$Q_{FB}^{\text{had}}$ measurements from the LEP and
SLD Collaborations~\cite{LEP-SLD}.}
\label{fig:stw_combined}
\end{center}
\end{figure}

To have a consistent SM definition and make our result comparable with previous measurements,
a LO {\sc pythia}
interpretation of the weak mixing angle with CTEQ6.6 PDF set~\cite{cteq66} is compared
to the predictions from a modified NLO {\sc resbos} with the same PDF set.
{\sc resbos} has a more sophisticated treatment of electroweak effects and uses different
values of effective weak mixing angle for leptons and up or down quarks~\cite{zgrad2}.
This study indicates that a $0.00008$ positive shift in $\stw$ for {\sc resbos} relative to LO
{\sc pythia} that changes the measured leptonic effective weak mixing angle to
$\effstw = 0.23147 \pm 0.00047$, with the same breakdown of uncertainties as above.
The comparison between our measurement and other experimental results is shown in Fig.~\ref{fig:stw_combined}.
Our measurement is consistent with the current world average.

In conclusion, we have measured the effective weak mixing angle $\effstw$ from the distribution of
the forward-backward charge
asymmetry $A_{FB}$ in the process $\DY$ at the Tevatron. This measurement, which supersedes
that reported in~\cite{d0-zee-5fb}, uses nearly twice the integrated luminosity and significantly
extends the electron acceptance.
The primary systematic uncertainty is reduced by introducing a new electron energy calibration method.
The result from 9.7 fb$^{-1}$ of integrated luminosity is $\effstw=0.23147 \pm 0.00047$,
This result is the most precise measurement from
light quark interactions, and is close to the precision of the world's best
measurements performed by the LEP and SLD Collaborations.

\section{Acknowledgements}
%

We thank the staffs at Fermilab and collaborating institutions,
and acknowledge support from the
Department of Energy and National Science Foundation (United States of America);
Alternative Energies and Atomic Energy Commission and
National Center for Scientific Research/National Institute of Nuclear and Particle Physics  (France);
Ministry of Education and Science of the Russian Federation, 
National Research Center ``Kurchatov Institute" of the Russian Federation, and 
Russian Foundation for Basic Research  (Russia);
National Council for the Development of Science and Technology and
Carlos Chagas Filho Foundation for the Support of Research in the State of Rio de Janeiro (Brazil);
Department of Atomic Energy and Department of Science and Technology (India);
Administrative Department of Science, Technology and Innovation (Colombia);
National Council of Science and Technology (Mexico);
National Research Foundation of Korea (Korea);
Foundation for Fundamental Research on Matter (The Netherlands);
Science and Technology Facilities Council and The Royal Society (United Kingdom);
Ministry of Education, Youth and Sports (Czech Republic);
Bundesministerium f\"{u}r Bildung und Forschung (Federal Ministry of Education and Research) and 
Deutsche Forschungsgemeinschaft (German Research Foundation) (Germany);
Science Foundation Ireland (Ireland);
Swedish Research Council (Sweden);
China Academy of Sciences and National Natural Science Foundation of China (China);
and
Ministry of Education and Science of Ukraine (Ukraine).


\begin{thebibliography}{99}
\bibitem{d0_coordinate}
D0 uses a cylindrical coordinate system with the $z$ axis
along the beam axis in the proton direction. Angles $\theta$ and $\phi$ are the polar and
azimuthal angles, respectively. Pseudorapidity is defined as
$\eta=-\ln [\tan(\theta/2)]$ where $\theta$ is measured with respect
to the interaction vertex. In the massless limit, $\eta$ is equivalent to
the rapidity $y=(1/2) \ln[(E+p_z)/(E-p_z)]$,
and $\eta_{\text{det}}$ is the pseudorapidity measured with respect to the
center of the detector.
\end{thebibliography}
\end{document}